\PassOptionsToPackage{dvipsnames,table}{xcolor} % Pass options FIRST
\documentclass[10pt, conference]{IEEEtran}
\IEEEoverridecommandlockouts
\usepackage{cite}
\usepackage{amsmath,amssymb,amsfonts}
\usepackage{algorithmic}
\usepackage{graphicx}
\usepackage{textcomp}
\usepackage{xcolor}
\def\BibTeX{{\rm B\kern-.05em{\sc i\kern-.025em b}\kern-.08em
    T\kern-.1667em\lower.7ex\hbox{E}\kern-.125emX}}

\usepackage{xcolor} % Load ONCE with all options
\usepackage{xspace}
\usepackage{listings}
\usepackage{subcaption}
\usepackage{comment}
\usepackage{tcolorbox}

\usepackage{tabularx}% http://ctan.org/pkg/tabularx
\usepackage{colortbl}% http://ctan.org/pkg/colortbl

\usepackage{url}
\usepackage{hyperref}

\usepackage{graphicx}

\usepackage{array}			% align vertically in tables
\usepackage{multirow}
\usepackage{tabularx}
\usepackage{booktabs} % Horizontal rules in tables
\usepackage{tikz}

\newcommand{\oapp}{CuRev\xspace}

\newcommand{\eg}{e.g.,~}							% exempli gratia (for the sake of example)
\newcommand{\ie}{i.e.,~}							% id est (that is)
\newcommand{\etal}{~et al.}					% et alia (and others)
\newcommand{\Fig}[1]{Figure~\ref{#1}}  			% choose Fig. or Figure, depending on the style
\newcommand{\Table}[1]{Table~\ref{#1}}	    % Table reference
\newcommand{\Sect}[1]{Section~\ref{#1}}	  % section name always with a capital S
\newcommand{\up}{\textcolor{ForestGreen}{\(^\uparrow\)}}
\newcommand{\down}{\textcolor{red!100!black}{\(^\downarrow\)}}

\begin{document}

% Harnessing LLMs for Curated Code Reviews
% Curating review comments for improved code review automation

\title{Harnessing Large Language Models for Curated Code Reviews\\
\thanks{The data is available at \url{https://zenodo.org/records/14812107}. The replication package is available at \url{https://github.com/OussamaSghaier/CuREV}.}
}

\author{\IEEEauthorblockN{Oussama Ben Sghaier}
\IEEEauthorblockA{\textit{Université de Montréal} \\
Montréal, Canada \\
oussama.ben.sghaier@umontreal.ca}
\and
\IEEEauthorblockN{Martin Weyssow}
\IEEEauthorblockA{\textit{Singapore Management University} \\
Singapore \\
mweyssow@smu.edu.sg}
\and
\IEEEauthorblockN{Houari Sahraoui}
\IEEEauthorblockA{\textit{Université de Montréal} \\
Montréal, Canada \\
sahraouh@iro.umontreal.ca}
}

\maketitle

\begin{abstract}
In code review, generating structured and relevant comments is crucial for identifying code issues and facilitating accurate code changes that ensure an efficient code review process. Well-crafted comments not only streamline the code review itself but are also essential for subsequent tasks like code refinement, where the code is modified to satisfy the input review comment. Although various AI-based approaches aimed to automate comment generation, their effectiveness remains limited by the quality of the training data. Existing code review datasets are often noisy and unrefined, posing limitations to the learning potential of AI models and hindering the automation process.

To address these challenges, we propose a curation pipeline designed to enhance the quality of the largest publicly available code review dataset. We begin by establishing an evaluation framework, incorporating specific criteria and categories to empirically study the initial quality of the dataset. Using a large language model (LLM)-driven approach, we then apply our curation pipeline to refine the dataset. A comparative analysis of the newly curated dataset, based on the same evaluation framework, demonstrates substantial improvements in the clarity and conciseness of the comments. Additionally, we assess the impact of the curated dataset on automating downstream tasks, specifically comment generation and code refinement. Our findings show that the curated dataset leads to enhanced model performance in generating more accurate comments. Curated comments are also more useful as they lead to more accurate code refinement.
\end{abstract}

\section{Introdution}
\label{sec:intro}

% General context of code review process
Code review is a critical component of the software development life cycle, aimed at identifying issues, detecting suboptimal code, and uncovering bugs \cite{mcintosh2014impact, mcintosh2016empirical}, while ensuring the overall quality and maintainability of the source code \cite{ackerman1989software, ackerman1984software, morales2015code}. This process typically involves a manual inspection of code by one or more developers, reviewing code written by their peers \cite{fagan2002design, bavota2015four}. The code review process consists of several key tasks, with the most essential being the identification and documentation of potential issues through review comments, the subsequent code refinement to resolve these concerns, and the quality assessment of the submitted code to decide if it should be accepted or needs further review.

% Put emphasis on the importance of 'comment generation' task 
Issue identification and description (\ie review comment generation) constitutes a foundational task in code review, focusing on the detection of defects or problems within the code. This phase is pivotal, as it involves not only identifying specific issues but also offers potential solutions for resolving them \cite{mcintosh2014impact, mcintosh2016empirical}. The significance of this task cannot be overstated, as subsequent stages of the review process heavily depend on its accuracy and thoroughness. Code refinement, for example, is an execution phase directly tied to insights gained from issue identification; developers revise and improve the code based on comments provided during this stage \cite{bacchelli2013expectations}. Thus, comment generation is essential to the entire code review process. Without precise execution of this task, the integrity and quality of later stages is compromised. As such, this task must be carried out meticulously, ensuring that issues and improvements are thoroughly examined and well-documented.

% Automation of comment generation
Automating comment generation has gained increasing attention, with various techniques proposed to improve accuracy and efficiency \cite{tufan2021towards, tufano2022using, li2022automating, li2022auger, ben2024improving}.
These methods typically use machine learning and natural language processing to analyze code and generate comments that identify defects and suggest fixes.
However, current systems still fall short of matching human reviewers in detecting nuanced issues and providing context-aware suggestions.
There is significant room for improvement across multiple facets of this task, from enhancing the quality of the existing datasets to developing more sophisticated generative language models.
Current models often struggle with producing comments that are both accurate and actionable, indicating a need for more refined approaches that can bridge the gap between human expertise and automated solutions.

% Challenges of automating comment generation
One key element to the success of deep learning models is the quality of training data. Language models cannot perform effectively if the training data is flawed or insufficient. Thus, high-quality data is foundational to the learning process, as models are only as good as the data they are trained on. In the case of code review, the datasets currently available \cite{li2022automating, sghaier2023multi}, while valuable, often suffer from several limitations. These datasets are frequently extracted in a raw form from repositories and are not curated or pre-processed to ensure their quality. As a result, they may contain various forms of noise, such as uncivil comments, irrelevant or poorly structured reviews, grammatical errors, or irrelevant comments. These issues not only hinder the learning process but can also lead to models learning undesirable patterns, such as generating irrelevant or incoherent feedback, thereby undermining the effectiveness of the automated task. Consequently, improving the quality of code review datasets is crucial for enhanced automation of comment generation and code refinement tasks.

%\begin{figure}[!t]
%    \centering
%    \includegraphics[width=\linewidth]{figures/eval_framework.pdf}
%    \caption{Overview of our evaluation framework.}
%    \vspace{-2em}
%    \label{fig:eval_framework}
%\end{figure}

In this paper, we propose an evaluation framework to categorize and assess the quality of review comments.
Fig.~\ref{fig:eval_framework} illustrates our evaluation framework consisting of (1) a categorization scheme to classify the type, nature, and civility of review comments and (2) scoring criteria to assess the overall quality of review comments based on relevance, clarity, and conciseness.
We apply our evaluation framework to the largest existing dataset of code reviews~\cite{li2022automating}.
Given the scale of the dataset, we employ a large language model (LLM) as a judge (i.e., LLM-as-a-Judge~\cite{zheng2023judging}) to automatically annotate samples with thoroughly designed prompts to ensure reliable and consistent annotations.
We show that $85.9\%$ of the samples are related to refactoring and bugfix, with a high proportion of prescriptive comments ($62.6\%$). 
Our analysis also reveals the presence of uncivil, lengthy, unclear, and irrelevant comments, highlighting the need to improve the dataset quality.

Subsequently, we introduce \oapp, a curated dataset for code review featuring a curation pipeline to improve the clarity, conciseness, and civility of review comments. 
Specifically, we filter out irrelevant comments from the original dataset and use \emph{Llama-3.1-70B} as LLM to reformulate review comments. 
This reformulation aims at maintaining the semantics of the original review comment while improving its form concerning the three scoring criteria. 
Our curated dataset features clearer, more concise, and 100\% civil review comments. 
To illustrate the actionability of \oapp, we conduct a comparative study with the original dataset by investigating their impact on the automation of comment generation and code refinement tasks.
We fine-tune \emph{DeepSeek-6.7B-Instruct}~\cite{deepseek-coder} using \oapp and the original dataset. 
Our results show a 46\% improvement in BLEU score for comment generation and a 22\% improvement in CodeBLEU for code refinement when fine-tuning the model using \oapp compared to the original dataset.

The main contributions of this paper are threefold: (1) the design of an evaluation framework to categorize and assess the quality of code review datasets, (2) the development of an automated curation pipeline to enhance dataset quality, and (3) a comparative study between our curated dataset, \oapp, and the original dataset, demonstrating the significant benefits of our curation pipeline in improving the effectiveness of LLMs for code review tasks.

\section{Methodology}
\label{sec:methodology}

In this section, we provide an overview of our methodology. Our goal is to answer the following research questions (RQs):
\begin{itemize}
    \item \textbf{RQ1:} \emph{What are the main characteristics and concerns in the code reviews dataset?}
    \item \textbf{RQ2:} \emph{How can our data curation pipeline improve the overall quality of the dataset, specifically in terms of reducing noise and irrelevant reviews?}
    \item \textbf{RQ3:} \emph{What is the impact of dataset curation on the performance of LLMs for automated comment generation?}
    \item \textbf{RQ4:} \emph{What is the impact of dataset curation on the comments' usefulness and LLMs performance for automated code refinement?}
\end{itemize}

\begin{figure*}[!h]
    \centering
    \includegraphics[width=0.9\linewidth]{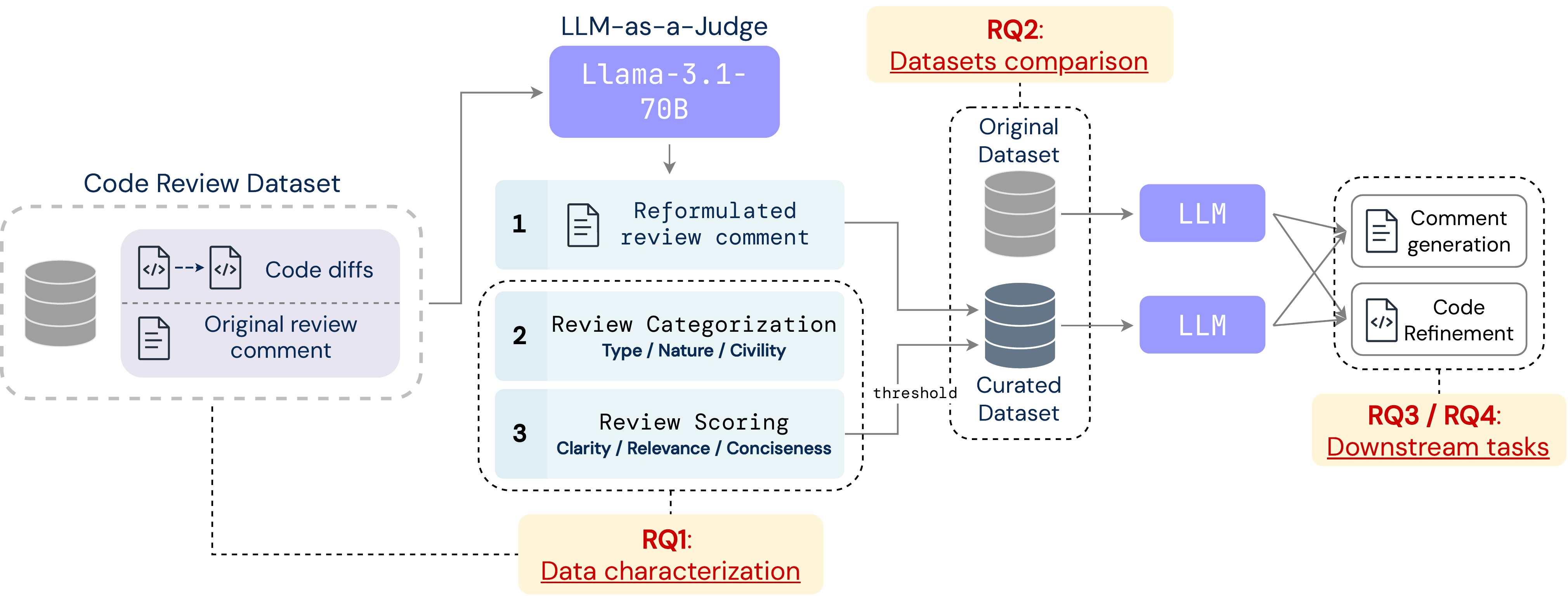}
    \caption{Overview of our methodology. We use a large code review dataset of samples comprising pre-commit and post-commit codes along with review comments. For each sample, we use LLM-as-a-Judge with Llama-3.1-70B to generate a reformulated review comment, a categorization of the review, and a score for the original review comment. Next, we use the reformulated review comments to create our curated dataset, while filtering out irrelevant samples. Finally, we compare the effectiveness of LLMs fine-tuned on the original and curated datasets on two downstream tasks: comment generation and code refinement.}
    \label{fig:methodology}
    \vspace{-.5em}
\end{figure*}

\begin{comment}
\begin{figure*}[!htbp]
    \centering
    \includegraphics[width=\textwidth]{figures/methodology.png}
    \caption{Overview of the proposed methodology.}
    \label{fig:methodology}
\end{figure*}
\end{comment}

\begin{figure}[!t]
    \centering
    \includegraphics[width=\linewidth]{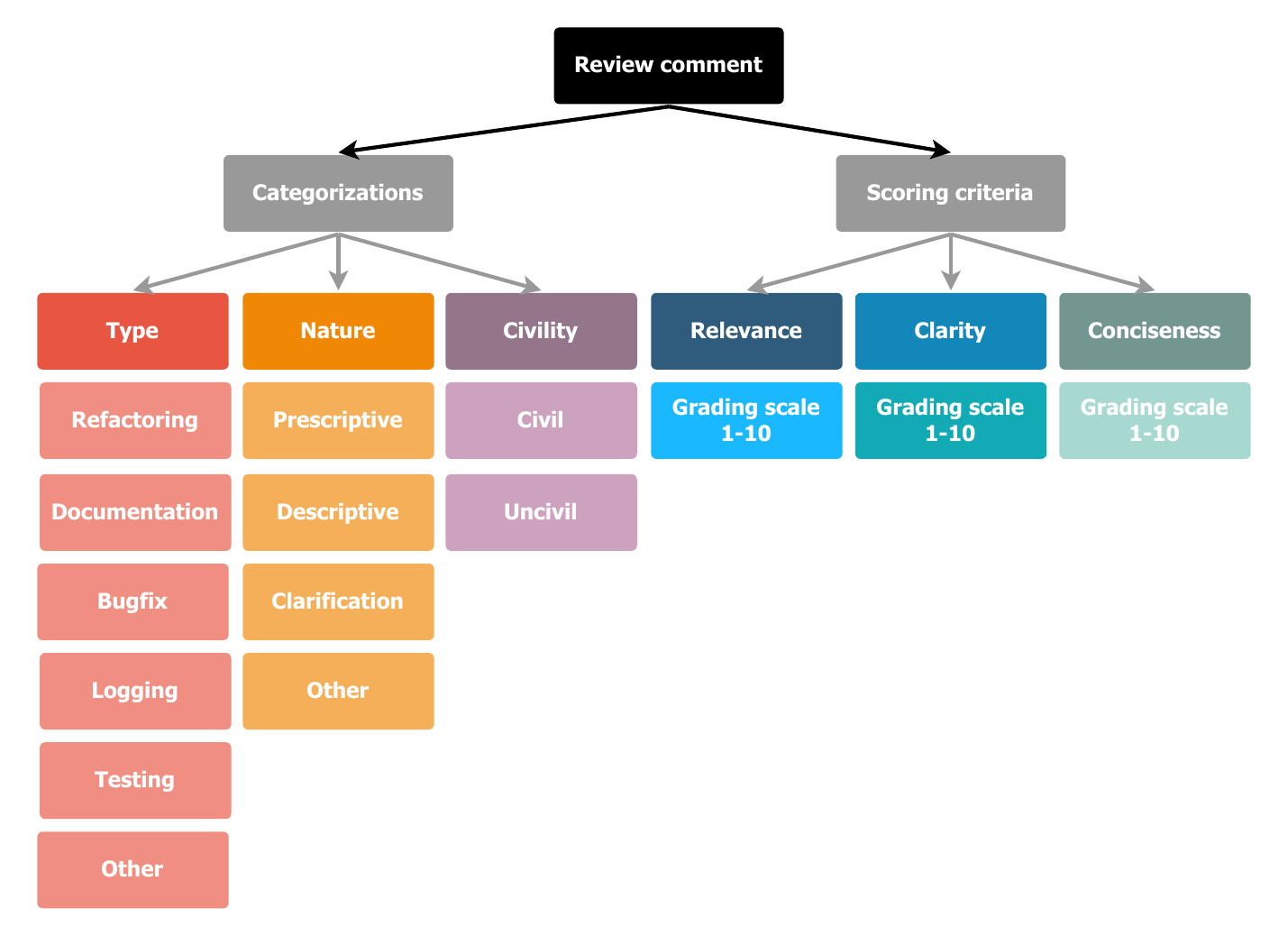}
    \caption{Overview of our evaluation framework.}
    % \vspace{-2em}
    \label{fig:eval_framework}
\end{figure}

\Fig{fig:methodology} illustrates the workflow of the proposed methodology. 
The process begins with selecting a large code review dataset \cite{li2022automating}, which forms the basis for subsequent analyses.
Then, we define an evaluation framework (see \Fig{fig:eval_framework}) to classify review comments according to multiple categories (\ie type, nature, civility) and to score them based on various criteria (\ie relevance, clarity, conciseness). We conduct a quality assessment and characterization of the dataset, utilizing an LLM, \emph{Llama-3.1-70B}, as an automated judge using the defined evaluation framework (\emph{RQ1}).

Subsequently, we propose a curation pipeline aimed at refining the review comments. This step involves employing \emph{Llama-3.1-70B} to reformulate the comments based on some criteria to enhance their quality. A comparative analysis of both the original and curated datasets is then performed using the defined evaluation framework, focusing on classification categories and scoring criteria (\emph{RQ2}).

The next phase consists of fine-tuning two separate language models, each trained on either the original or curated dataset, to automate the task of comment generation. The performance of these models is evaluated and compared to determine which dataset better facilitates the learning process for generating high-quality review comments (\emph{RQ3}).

Finally, we assess the usefulness of the review comments for the subsequent task (\ie code refinement). We use two versions of the same language model, each provided with one version of the review comments (\ie original or curated), and compare the accuracy of the generated code. This comparison allows us to evaluate the impact of the curated versus original review comments on the correctness and effectiveness of automated code refinement (\emph{RQ4}).

\section{Quality Assessment and Characterization of Existing Code Review Dataset}
\label{sec:initdata}

In this section, we present the evaluation framework, detail its application in assessing the existing code review dataset, and then present the results.

\subsection{Dataset Selection}
We employ the code review dataset introduced in \cite{li2022automating}, which is the largest publicly available dataset for code reviews. This dataset has been widely adopted in several works \cite{li2022automating, ben2024improving, sghaier2023unity, lu2023llama} for automating code review tasks.
The dataset is multilingual, covering nine programming languages, and consists of $176,613$ samples. \Table{tab:dataset_stats} shows the distribution of the dataset across the different programming languages.
The dataset includes features such as the code changes submitted for review and the review comment provided by the reviewer assigned to the pull request. 

% The dataset includes the following key features:
% \begin{itemize} 
% \item Code changes submitted for review 
% \item The full original code file associated with the review 
% \item The review comment provided by the reviewer assigned to the pull request 
% \item Code edits that implement the changes requested in the review comment 
% \item GitHub repository information 
% \item Programming language 
% \end{itemize}

\begin{table}[!t]
  \centering
  \caption{Dataset distribution over programming languages.}
  \label{tab:dataset_stats}
  \begin{tabular}{{>{\centering\arraybackslash}p{1.5cm}}*{1}{r}}
    \toprule
    \textbf{Programming Language} & \textbf{\# Samples} \\
    \midrule
    PHP & $9,984$ \\
    Ruby & $6,713$ \\
    C\# & $17,085$ \\
    C & $4,108$ \\
    Java & $35,671$ \\
    Python & $36,382$ \\
    C++ & $15,944$ \\
    Go & $36,123$ \\
    JS & $14,603$ \\
    \midrule
    \textbf{Total} & $176,613$\\
    \bottomrule
  \end{tabular}
  \vspace{-1em}
\end{table}

\subsection{Evaluation Framework}

In this section, we introduce the evaluation framework used to assess the quality and characteristics of the code reviews in existing datasets. 
As shown in \Fig{fig:eval_framework}, this framework consists of three key categorizations, along with a scoring system to evaluate the clarity, relevance, and conciseness of each review comment. These categories and criteria aim to provide a comprehensive analysis of the reviews and enable us to identify areas for improvement in the dataset.

\Table{tab:categories} presents the categorization framework for code reviews, which is used to classify each review comment based on three key aspects: \emph{Type}, \emph{Nature}, and \emph{Civility}. The \emph{Type} category identifies the primary focus of the review, such as refactoring, bugfixes, or documentation \cite{tufano2024code}. The \emph{Nature} category assesses the intent behind the comment, categorizing it as prescriptive, descriptive, or clarifying. Finally, the \emph{Civility} category evaluates the tone of the comment; whether it is civil or uncivil \cite{rahman2024words}. 
Note that \emph{Type} and \emph{Nature} categories are multi-labeled. That is, a single review comment may address multiple types of issues (\eg documentation and testing) or have a mixed nature (\eg descriptive and requests clarifications).
This categorization provides a structured approach for systematically evaluating the dataset, offering a detailed breakdown of the content and intent of review comments.

\begin{table}[!t]
  \centering
  \caption{Categorization framework for reviews comments.}
  % \vspace{-.5em}
  \label{tab:categories}
  \begin{tabularx}{\linewidth}{llX}
    \toprule
    \textbf{Category} & \textbf{Subcategory} & \textbf{Description} \\
    \midrule
    \textbf{Type} & Refactoring & Suggestions to improve code structure \\
                  & Bugfix & Identifies and suggests fixes for bugs \\
                  & Testing & Comments related to test cases \\
                  & Logging & Suggestions for logging practices \\
                  & Documentation & Recommendations for documentation changes \\
                  & Other & Any other type of comment \\
    \midrule
    \textbf{Nature} & Prescriptive & Provides specific actions or recommendations \\
                    & Descriptive & Describes a situation without suggesting changes \\
                    & Clarification & Requests or provides clarification \\
                    & Other & Any other nature of comment \\
    \midrule
    \textbf{Civility} & Civil & Respectful and professional tone \\
                      & Uncivil & Disrespectful or inappropriate tone \\
    \bottomrule
  \end{tabularx}
  \vspace{-1em}
\end{table}

\Table{tab:criteria} outlines the scoring criteria used to assess the quality of reviews in three dimensions: \emph{clarity}, \emph{relevance}, and \emph{conciseness} \cite{rani2023decade, haouari2011good}. Each criterion is scored on a scale from 1 to 10. \emph{Clarity} measures how effectively the comment communicates its message, with higher scores reflecting clearer communication. \emph{Relevance} evaluates how pertinent the comment is to the code change, and \emph{Conciseness} assesses whether the comment is brief and to the point without unnecessary details. These criteria provide a structured way to evaluate the quality and effectiveness of each comment, ensuring a comprehensive assessment of the dataset.

\begin{table}[!t]
  \centering
  \caption{Scoring criteria for review comments.}
  % \vspace{-.5em}
  \label{tab:criteria}
  \begin{tabularx}{\linewidth}{lX}
    \toprule
    \textbf{Criterion} & \textbf{Description} \\
    \midrule
    \textbf{Clarity (1--10)} & Assesses how clearly the review comment communicates its message. A score of 1 indicates very unclear, and 10 indicates very clear communication. \\
    \textbf{Relevance (1--10)} & Evaluates the extent to which the review comment is pertinent to the code change. A score of 1 means the comment is completely irrelevant, while a score of 10 means it is highly relevant. \\
    \textbf{Conciseness (1--10)} & Measures the brevity and efficiency of the review comment, ensuring it conveys the necessary information without unnecessary elaboration. A score of 1 indicates too verbose, and 10 indicates concise and to the point. \\
    \bottomrule
  \end{tabularx}
  \vspace{-1em}
\end{table}

This framework enables a structured analysis of the reviews, allowing us to assess their quality in a granular manner and providing the foundation for identifying trends, patterns, and areas for improvement within the dataset.

\subsection{LLM-as-a-Judge}

To implement the evaluation framework on the code review dataset, we leverage an LLM, specifically \emph{Llama-3.1-70B-Instruct}, to act as annotator for the various categories and criteria. Due to the substantial size of the dataset (176,613 samples), manual evaluation is infeasible. Drawing from findings in the literature~\cite{zheng2023judging, zhuo2023ice, chang2024survey}, we assume that a highly capable LLM can serve as a reliable substitute for human evaluators, accurately assessing review comments across the defined framework.
This claim is supported by previous research demonstrating that highly capable LLMs closely align with human judgments, achieving agreement rates comparable to human-to-human agreement~\cite{zheng2024judging, li2023alpacaeval}. Furthermore, LLMs have been recently exploited as judges for different software engineering tasks~\cite{zhuo2023ice, weyssow2024codeultrafeedback}, thereby justifying their use as annotators in this work.

For conciseness, we show an excerpt of the prompt \footnote{The full prompt is available in the \href{https://github.com/AI4CodeReview/CuREV}{replication package}.} used to evaluate review comments in \Table{tab:eval_promp}.
Initially, the LLM is instructed to generate what it considers an ideal review comment based on the provided code changes. 
Following this, the LLM is asked to evaluate the given review comment according to the different categories and criteria. 
The generated review serves as an implicit reference during this evaluation. 
This process is repeated across the entire code review dataset.

% Prompt design
The prompt was chosen through multiple refinement iterations, following a trial-and-error approach based on observations. Initially, we designed a preliminary prompt and tested it on a curated set of manually selected examples. Based on the identified issues, we refined the prompt to enhance its effectiveness. This iterative process was informed by techniques from the literature that demonstrated effectiveness in similar contexts \cite{weyssow2024codeultrafeedback}. 
For instance, one challenge encountered was the tendency of LLM to generate unstructured and verbose outputs in inconsistent formats, which hindered automated parsing and extraction of relevant information. To mitigate this issue, the prompt was augmented with explicit examples of the desired output format, thereby encouraging the LLM to produce more structured and consistent results.
Another observed limitation was the propensity of LLM to assign high scores (\eg 9 or 10) across all evaluation criteria, resulting in a lack of meaningful differentiation. To address this, detailed descriptions of each criterion were incorporated into the prompt, and the LLM was instructed to generate an exemplary review as a reference. This adjustment facilitated a more balanced and discriminative grading distribution. Additionally, we asked the LLM to provide a rationale for each evaluation, effectively introducing an implicit chain-of-thought process. This approach not only improved the explainability and justification of the assigned grades but also enhanced the overall reliability and quality of the evaluations.

\begin{table}[!t]
\centering
\caption{Excerpt of the prompt for comments evaluation.}
\vspace{-.5em}
\label{tab:eval_promp}
\begin{tabularx}{\linewidth}{X}
\toprule
\textbf{\#\#\# Code review comment generation}

 Generate a review comment that you consider perfect for the code change without considering the given input comment. A review comment should highlight the main issues, improvements, or suggestions for the code changes. The generated review comment should be concise, relevant, clear, useful, and complete. \\
\vspace{.3em}
\textbf{\#\#\# Code review comment assessment}

Then, evaluate and categorize only the given review comment, written by a reviewer, based on the below criteria.
You can use the generated review comment as a reference to evaluate the given review comment.
Note that multiple labels are allowed for the categories "Type" and "Nature". \\
\\
\textsc{\textbf{1. Type:}} Categorize the review according to the type of issue it addresses: Refactoring, Bugfix, Testing, Logging, Documentation, Other.
\vspace{.3em}

\textsc{\textbf{2. Nature:}} Specify the nature of the review according to these categories:\\
- \textit{Descriptive:} describe what the reviewer observes without explicitly suggesting specific actions.\\
- \textit{Prescriptive:} suggest or request specific actions on the code.\\
- \textit{Clarification:} request explanation or further information to better understand the code changes.\\
- \textit{Other:} for comments that do not fit the previous categories.

\hspace{0.5\linewidth}\vdots

\textsc{\textbf{4. Conciseness:}} Assess how effectively the comment conveys its message using the fewest necessary words while remaining fully informative. A concise comment should be completely brief but informative, avoiding unnecessary details, repetition, or verbosity. Use a 1-to-10 rating scale.

\hspace{0.5\linewidth}\vdots

\textbf{\#\#\# Given review comment}\\
\{\textit{review\_comment\}}\\

\textbf{\#\#\# Code changes}\\
\{\textit{code\_diff\}}\\
\bottomrule
\end{tabularx}
\vspace{-1em}
\end{table}

\subsection{Sanity Check}
To ensure the reliability of the judgments made by \emph{Llama-3.1-70B-Instruct}, we conducted a sanity check involving human assessments. A random sample of $100$ review comments was manually evaluated by two authors according to the evaluation schema defined in \Fig{fig:eval_framework}. This dual-author assessment aimed to mitigate individual biases and provide a robust baseline for comparison. Conflicts between the annotators were carefully resolved through discussion to reach a consensus on each sample's evaluation.

% This manual assessment provides a human baseline against which we can compare the LLM judging performance. To measure the agreement between LLM and human judgments, we used \emph{Cohen's kappa}, a statistical measure that accounts for the degree of agreement beyond chance \cite{mchugh2012interrater}. Our analysis resulted in a Cohen's kappa value of \obs{XX}, indicating substantial agreement between the LLM's assessments and human judgments.

This manual assessment provides a human baseline against which we can compare the LLM judging performance. To measure the agreement between LLM and human judgments, we used \emph{Cohen's kappa}, a statistical measure that accounts for the degree of agreement beyond chance \cite{mchugh2012interrater}. Our analysis showed \emph{perfect agreement} for the \emph{civility} category ($1$), \emph{near-perfect agreement} for \emph{type} ($0.88$) and \emph{nature} ($0.82$) categories. For criteria, we observed \emph{near-perfect agreement} for \emph{relevance} ($0.85$), \emph{substantial agreement} for \emph{Conciseness} ($0.76$) and \emph{Clarity} ($0.64$). These values provide strong evidence of the LLM capacity to make reliable judgments.

This outcome supports existing literature suggesting that LLMs possess the capability to evaluate review comments with accuracy comparable to human reviewers \cite{zheng2024judging, li2023alpacaeval}. Furthermore, it underscores the reliability of LLMs in the specific context of judging review comments.

\subsection{Categorization Results}

\begin{figure*}
    \centering
    \includegraphics[width=.9\linewidth]{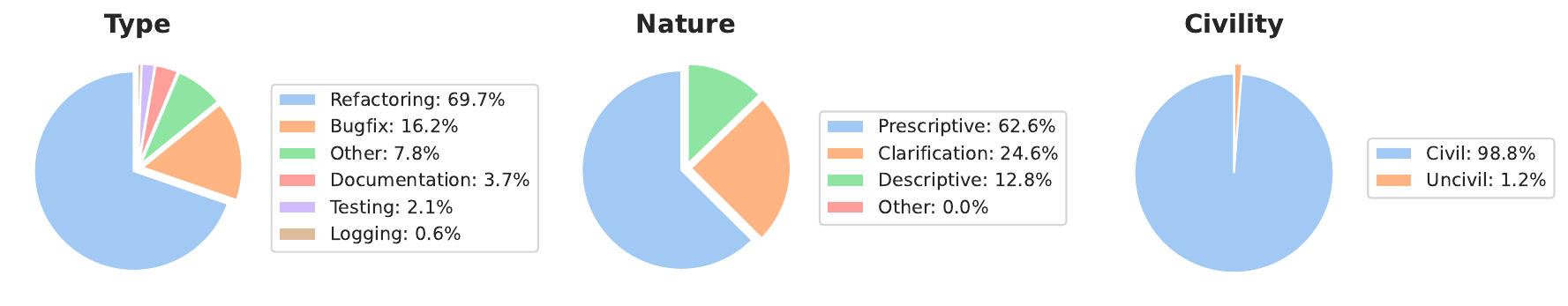}
    % \vspace{-1em}
    \caption{Distribution of the different categories across the original dataset.}
    \label{fig:initinal_categ_dist}
    \vspace{-1em}
\end{figure*}

\Fig{fig:initinal_categ_dist} presents the results of the different categorizations according to the \emph{type}, \emph{nature}, and \emph{civility} of the comments. The findings provide valuable insights into the original dataset characteristics and highlight potential areas of improvements.

\paragraph*{\textbf{Type of review comments}}

The majority of the review comments in the dataset fall under the \emph{Refactoring} category, comprising $80.07\%$ of the total. This indicates that much of the feedback is centered on restructuring the code to enhance its quality and performance. The next most significant category is \emph{Bugfix} at $18.60\%$, underscoring the importance of addressing functional issues within the code to avoid bugs.

Other subcategories, such as \emph{Documentation} ($4.21\%$), \emph{Testing} ($2.42\%$), and \emph{Logging} ($0.65\%$), are relatively less frequent. This suggests that these areas receive less attention in code review feedback, potentially due to implicit understanding among developers or prioritization of other aspects of the code. The \emph{Other} category, which makes up $8.97\%$ of the comments, encompasses miscellaneous feedback types that do not fall into the primary classifications, \eg security.

The focus on \emph{Refactoring} and \emph{Bugfix} comments align with the primary objectives of code review, \ie preserving code quality and preventing bugs. However, the lower representation of \emph{Documentation}, \emph{Testing}, and \emph{Logging} comments suggests potential gaps in the comprehensiveness of code review practices. Addressing this imbalance presents an opportunity for enhancing training data to support more well-rounded feedback generation. This finding can guide researchers in balancing their datasets according to the \emph{type} of review comment, ultimately leading to more varied and comprehensive model-generated feedback.

\paragraph*{\textbf{Nature of review comments}}

Review comments categorized as \emph{Prescriptive}, which provide direct guidance or specific instructions, dominate the dataset at $62.6\%$. This reflects a strong focus on actionable feedback, aiding developers in making precise code changes. \emph{Clarification} comments make up $24.6\%$, indicating a notable effort to ensure code understanding among team members. \emph{Descriptive} comments, which explain aspects of the code without providing explicit recommendations, account for $12.8\%$.
The \emph{Other} category is negligible at $0.01\%$, showing that most comments fit well into defined subcategories.

The prevalence of \emph{prescreptive} comments highlights a code review culture geared towards providing direct solutions and explicit feedback. The significant presence of \emph{Clarification} comments suggests that maintaining clear communication and understanding is a priority during the review process.

\paragraph*{\textbf{Civility of review comments}}

Most comments in the dataset are categorized as \emph{Civil}, making up $98.77\%$ of the total. This suggests that the code review process is generally conducted professionally and constructively. However, there is still a portion ($1.23\%$) of \emph{Uncivil} comments present in the dataset. These comments often contain harsh or inappropriate language that could negatively influence a model learning process.
Training on such comments risks introducing undesirable patterns, potentially leading to the model generating inappropriate language in downstream tasks.

To mitigate this, it is crucial to either curate these reviews to remove harsh language or exclude them altogether to prevent models from learning from these undesirable examples and reinforcing negative behavior.

\subsection{Scoring Criteria Results}

\begin{figure}
    \centering
    \includegraphics[width=.8\linewidth]{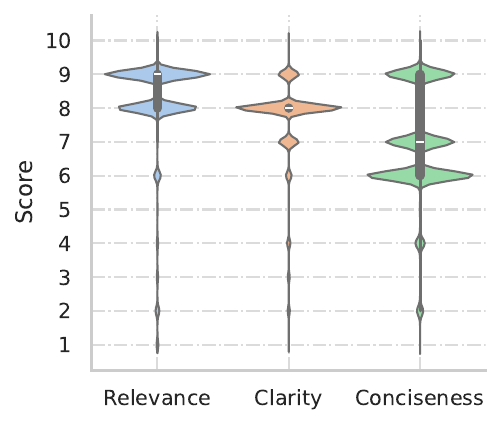}
    % \vspace{-.5em}
    \caption{Distribution of scoring criteria on the original dataset.}
    \label{fig:init_scoring_dist}
    \vspace{-1em}
\end{figure}

\Fig{fig:init_scoring_dist} depicts scoring criteria distribution across the original dataset. \Table{tab:init_categories_distribution} provides a summary of the average values for each scoring criterion within the different categories.

\begin{table}[!t]
  \centering
  \caption{Average values of the scoring criteria per category across the original dataset.}
  \label{tab:init_categories_distribution}
  \begin{tabular}{>{\centering\arraybackslash}p{1cm} >{\centering\arraybackslash}p{1.5cm}*{3}{c}}
    \toprule
    \textbf{Category} & \textbf{Subcategory} & \textbf{Relevance} & \textbf{Clarity} & \textbf{Conciseness} \\
    \midrule
    \multirow{6}{*}{\textbf{Type}} & Refactoring & $8.32$ & $7.79$ & $6.99$ \\
    & Bugfix & $8.53$ & $7.74$ & $6.84$ \\
    & Testing & $8.42$ & $7.92$ & $6.97$ \\
    & Logging & $8.43$ & $7.84$ & $6.85$ \\
    & Documentation & $8.33$ & $7.62$ & $6.72$ \\
    & Other & $7.02$ & $6.86$ & $5.90$ \\
    \midrule
    \multirow{4}{*}{\textbf{Nature}} & Descriptive & $7.14$ & $6.63$ & $5.61$ \\
    & Prescriptive & $8.52$ & $7.95$ & $7.19$ \\
    & Clarification & $8.29$ & $7.57$ & $6.66$ \\
    & Other & $4.24$ & $4.40$ & $4.12$ \\
    \midrule
    \multirow{2}{*}{\textbf{Civility}} & Civil & $8.26$ & $7.75$ & $6.93$ \\
    & Uncivil & $5.60$ & $4.34$ & $4.34$ \\
    \midrule
    \multirow{1}{*}{\textbf{Average}} & -- & $8.23$ & $6.89$ & $7.71$ \\
    \bottomrule
  \end{tabular}
  \vspace{-1em}
\end{table}

\paragraph*{\textbf{Relevance of review comments}}
The highest relevance scores are found in the \emph{Bugfix} ($8.53$) and \emph{Prescriptive} ($8.52$) subcategories. The lowest relevance scores are observed in the \emph{Other} subcategories of both the \emph{Type} ($7.02$) and \emph{Nature} ($4.24$) categories, as well as in \emph{Uncivil} comments ($5.60$).
The low relevance of these subcategories indicates that such comments may often lack the focus or constructive value necessary for high-quality code reviews. 

The relevance distribution, illustrated in \Fig{fig:init_scoring_dist}, indicates that while the majority of comments are relevant, a portion of irrelevant comments persists, diminishing the overall quality of the dataset. These irrelevant comments can negatively influence language models' performance by introducing unhelpful patterns. 
This study can support the filtering of irrelevant comments to ensure a high-quality dataset, as maintaining data relevance is crucial for effective model training and downstream task performance.

\paragraph*{\textbf{Clarity of review comments}}

Clarity scores across the dataset are varied, with the highest scores found in the \emph{Prescriptive} $(7.95$), \emph{Testing} ($7.92$), and \emph{Refactoring} ($7.79$) subcategories, while the lowest clarity scores appear in the \emph{Other} subcategories ($4.40$) and \emph{Uncivil} comments ($4.34$).

The data indicates that \emph{Prescriptive} comments, which provide direct and specific guidance, are not only relevant but also clear, making them highly effective in guiding developers. High clarity scores in the \emph{Testing} and \emph{Refactoring} categories further suggest that comments in these areas are typically well-understood. In contrast, comments categorized as \emph{Other} and \emph{Uncivil} exhibit significantly lower clarity, which can lead to misunderstandings and inefficiencies during the code review process. The low clarity score (average=$4.89$) underscores the importance of enhancing the clarity of review comments to improve dataset quality, ultimately supporting the training of language models to produce clear and precise feedback that developers can easily comprehend and act upon.

\paragraph*{\textbf{Conciseness of review comments}}

The analysis of conciseness scores reveals that the highest average scores are found in the \emph{Prescriptive} ($7.19$), \emph{Refactoring} ($6.99$), and \emph{Testing} ($6.97$) subcategories. The lowest conciseness scores are observed in \emph{Other} subcategories for both \emph{Type} ($5.90$) and \emph{Nature} ($4.12$), as well as in \emph{Uncivil} comments ($4.34$).

The high conciseness scores in \emph{Prescriptive} comments indicate that these types of feedback tend to be not only clear and relevant but also succinct, contributing to more efficient communication. 
The lower conciseness scores suggest that these comments may be more verbose or include unnecessary language, reducing their effectiveness.

As shown in \Fig{fig:init_scoring_dist}, a significant portion of the dataset consists of comments that lack conciseness. This suggests an opportunity for enhancing the quality of the dataset by refining review comments to be more succinct, removing unnecessary elements, and conveying the message more efficiently. This improvement would enable language models to focus on the core content of the comments, emphasizing essential information over extraneous details.

\begin{center}
\begin{tcolorbox}[colframe=teal!75!black, colback=teal!5!white, title=\textbf{Answer to RQ1}]
The code reviews dataset is characterized by a strong focus on \emph{refactoring} ($80.07\%$) and \emph{bugfix} ($18.60\%$). Most of the comments are \emph{prescriptive} ($62.6\%$) providing direct and actionable suggestions to developers.
However, the presence of uncivil, lengthy, unclear, and irrelevant comments highlights areas for improvement to enhance dataset quality.
\end{tcolorbox}
\end{center}
\section{CuRev: A Curated Dataset for Code Review}
\label{sec:curdata}

\subsection{Experimental Setup}
Based on our previous experiment findings, we developed a curation pipeline aimed at enhancing the overall quality of the code review dataset. This pipeline comprises two primary steps: filtering out irrelevant comments and reformulating comments to improve clarity, conciseness, and civility.

\paragraph*{\textbf{Filtering irrelevant review comments}}
The first step involves identifying and removing irrelevant review comments. While certain criteria, such as civility, clarity, and conciseness, can be improved through rephrasing, relevance is inherently static and tied to the core idea of the comment. A relevant or irrelevant comment retains this classification regardless of its formulation, as it addresses the same issue or improvement.

To implement this step, we carefully examined the relevance scores from the initial evaluation and set a threshold of $4$. Any comment with a relevance score below this threshold was considered irrelevant and subsequently filtered out. This filtering process resulted in the elimination of $5,895$ samples, leaving a curated dataset of size $170,718$.

These are three examples of review comments that were filtered out due to their low relevance scores ($<4$):
\begin{quote}
\textbf{Example 1}: \texttt{Need some edit here?}

\textbf{Example 2}: \texttt{Same here etc :)}

\textbf{Example 3}: \texttt{This is gross}
\end{quote}
While human reviewers might understand these comments in real-life contexts—where much communication occurs verbally and directly—they remain irrelevant and unhelpful for a model to learn from, as they lack useful information.

\paragraph*{\textbf{Review comments reformulation}}
The second step focuses on reformulating review comments to enhance the criteria of civility, clarity, and conciseness. For this task, we utilized an LLM, \emph{Llama-3.1-70B}, providing specific instructions to guide the reformulation process. The reformulation task was designed to maintain the original intent and content of each comment while improving its form and presentation. \Table{tab:reform_prompt} shows an excerpt of the prompt used for this task.

\begin{table}[!t]
\centering
\caption{Excerpt of the prompt for comment reformulation.}
% \vspace{-.5em}
\label{tab:reform_prompt}
\begin{tabularx}{\linewidth}{X}
\toprule
\textbf{\#\#\# Review comment reformulation}

Your task is to reformulate and improve the given review comment by making it civil, more clear, and more concise without changing its core message or intent. 
The reformulated comment should respect the following guidelines:\\
\\
\textsc{\textbf{1. Conciseness:}} The comment should convey its message in the fewest words necessary while still being informative. Eliminate redundancy and irrelevant details.
\vspace{.5em}

\textsc{\textbf{2. Clarity:}} Ensure the comment is straightforward, well-structured, and grammatically correct, making the feedback easy to understand without any ambiguity.
\vspace{.5em}

\textsc{\textbf{3. Civility:}} Keep the comment respectful, professional, and constructive, avoiding any harsh or inappropriate language.\\

\hspace{0.5\linewidth}\vdots

\textbf{\#\#\# Given review comment}\\
\{\textit{review\_comment\}}\\

\textbf{\#\#\# Code changes}\\
\{\textit{code\_diff\}}\\
\bottomrule
\end{tabularx}
\vspace{-1em}
\end{table}

The reformulation guidelines, employed in the prompt, emphasize three key criteria: conciseness, clarity, and civility. The reformulated comment should convey its message in the fewest words necessary while remaining informative, removing any redundancy or irrelevant details (\emph{conciseness}). It should also be well-structured, straightforward, and free of ambiguity to ensure ease of understanding (\emph{clarity}). Additionally, the comment must maintain a respectful and professional tone, providing constructive feedback without using harsh or inappropriate language (\emph{civility}).

% Handling non-english words
The code review dataset used is exclusively composed of English comments, though some comments may include non-English words depending on the context. A powerful and multilingual LLM, such as Llama-3.1-70B, can appropriately handle these cases effectively. This was supported by findings in the literature, which indicate that LLMs, despite being predominantly trained on large amounts of English data, can still manage non-English languages to a reasonable extent \cite{terryn2024exploratory}.

In the final step, we re-evaluated the curated review comments using the same evaluation schema as was applied to the original dataset. Relevance was omitted from this re-evaluation, as it does not change with reformulation; it is dependent on the content of the comment rather than its form. This re-evaluation allowed for a direct comparison between the original and curated review comments, assessing variations in clarity, conciseness, and civility.

\subsection{Results}

The analysis of the scoring criteria across the curated dataset, as shown in \Table{tab:cur_scoring_dist}, reveals improvements in both clarity and conciseness. The average clarity score increased to $8.96$, representing a substantial improvement of $2.07$. Conciseness also saw an increase, reaching an average score of $8.05$, which marks a smaller improvement of $0.34$.

The important improvement in clarity indicates that curated comments are more structured, eliminating noise and unnecessary words while using grammatically correct vocabulary and well-formed sentences, making them clearer and easier to understand. However, the modest improvement in conciseness can be attributed to the inherently verbose nature of LLMs, which tend to produce more elaborate text by default.

Overall, the enhancements in clarity and conciseness suggest that the curation pipeline successfully refined the dataset by producing review comments that are more succinct and easier to understand. These improvements likely contribute to better data quality, facilitating more effective learning by language models.

\begin{table}[!t]
  \centering
  \caption{Evolution of the scoring criteria per category across the curated dataset.}
  \label{tab:cur_scoring_dist}
  \begin{tabular}{>{\centering\arraybackslash}p{1cm} >{\centering\arraybackslash}p{1.5cm}*{2}{c}}
    \toprule
    \textbf{Category} & \textbf{Subcategory} & \textbf{Clarity} & \textbf{Conciseness} \\
    \midrule
    \multirow{6}{*}{\textbf{Type}} & Refactoring & $8.95$ (\up $0.63$) & $8.06$ (\up $1.07$) \\
    & Bugfix & $8.98$ (\up $0.45$) & $8.03$ (\up $1.19$) \\
    & Testing & $8.98$ (\up $0.56$) & $8.03$ (\up $1.06$) \\
    & Logging & $8.97$ (\up $0.54$) & $8.02$ (\up $1.17$) \\
    & Documentation & $8.98$ (\up $0.65$) & $8.02$ (\up $1.30$) \\
    & Other & $8.95$ (\up $1.93$) & $8.05$ (\up $2.15$) \\
    \midrule
    \multirow{4}{*}{\textbf{Nature}} & Descriptive & $8.76$ (\up $1.62$) &  $8.02$ (\up $2.41$)\\
    & Prescriptive & $8.96$ (\up $0.44$) & $8.05$ (\up $0.86$) \\
    & Clarification & $8.96$ (\up $0.67$) & $8.03$ (\up $1.37$) \\
    & Other & $9.00$ (\up $4.76$) & $8.00$ (\up $3.88$) \\
    \midrule
    \multirow{2}{*}{\textbf{Civility}} & Civil & $8.96$ (\up $0.70$) & $8.05$ (\up $1.12$) \\
    & Uncivil & -- & -- \\
    \midrule
    \multirow{1}{*}{\textbf{Average}} & -- & $8.96$ (\up $2.07$)  & $8.05$ (\up $0.34$) \\
    \bottomrule
  \end{tabular}
  \vspace{-1em}
\end{table}

\Table{tab:cur_categories_dist} presents the evolution of comment categorizations in the curated dataset compared to the original one. In the \emph{nature} category, the percentage of \emph{prescriptive} comments increased to $90.20\%$ (vs. $62.6\%$ in the original dataset) and the percentage of descriptive comments decreased significantly to $0.95\%$, demonstrating a strong shift towards more directive, explicit, and actionable guidance to developers. This could be explained by the fact that prescriptive comments might be clearer compared to descriptive comments.

In the \emph{civility} category, all comments were marked as \emph{civil}, achieving a perfect score ($100\%$) and eliminating any \emph{Uncivil} comments. This indicates that the curation process effectively addressed concerns related to inappropriate or harsh language, creating a more professional and constructive dataset.

\begin{table}[!htbp]
  \centering
  \caption{Categories statistics in the curated dataset.}
  \label{tab:cur_categories_dist}
  \begin{tabular}{>{\centering\arraybackslash}p{1.2cm} >{\centering\arraybackslash}p{2cm}*{2}{r}}
    \toprule
    \textbf{Category} & \textbf{Subcategory} & \textbf{Count} & \textbf{Percentage}\\
    \midrule
    \multirow{4}{*}{\textbf{Nature}} & Descriptive & $1,674$ & $0.95$\down \\
    & Prescriptive & $159,306$ & $90.20$\up \\
    & Clarification & $29,586$ & $16.75$\down \\
    & Other & $17,491$ & $9.90$\up \\
    \midrule
    \multirow{2}{*}{\textbf{Civility}} & Civil & $176,613$ & $100$\up \\
    & Uncivil & $0$ & $0$\down \\
    \bottomrule
  \end{tabular}
  \vspace{-1.5em}
\end{table}

\Table{tab:reform_examples} presents examples from the dataset illustrating the original and reformulated review comments. The original comments contain inappropriate language and lack professionalism, which could impede effective communication and contribute to an unconstructive code review environment. The reformulated versions demonstrate significant improvements in clarity and civility. For instance, the original comment, “\texttt{you need to use list\_delete here, list\_free doesn't do shit...}” was revised to a more professional and instructive tone: “\texttt{Consider using list\_delete instead of list\_free to properly clean up the nh\_list ...}”. Similarly, the second example shows a shift from an uncivil and vague comment to one that is respectful and precise. These examples reflect the impact of the curation pipeline in producing comments that are clearer, more actionable, and maintain a civil tone, making them more suitable for use in automated code review tools.

\begin{table*}[!htbp]
\centering
\caption{Examples of original and reformulated review comments from \oapp.}
\label{tab:reform_examples}
\begin{tabularx}{\textwidth}{XX}
\toprule
\multicolumn{1}{c}{\textbf{Original Review Comment}} & \multicolumn{1}{c}{\textbf{Reformulated Review Comment}} \\
\midrule

\texttt{you need to use `list\_delete` here, `list\_free` doesn't do shit... (honestly `list\_free` shouldn't be in the API to begin with)} &
\texttt{Consider using `list\_delete` instead of `list\_free` to properly clean up the `nh\_list` in `ospf6\_route\_delete`. The `list\_free` function does not handle the deletion of the list's elements, which is necessary in this case.} \\
\midrule

% \texttt{wtf? `r\_buf\_buffer` is now returning a new buffer with all the bytes? this is gonna be a huge pain for big files} &
% \texttt{Consider the performance implications of `r\_buf\_buffer` now returning a new buffer with all bytes, especially for large files. This change may introduce significant memory overhead. Could you provide insight into the reasoning behind this change and potential optimizations for handling big files?} \\
% \midrule

\texttt{mmmm welcome to the fucking world of TS ids... -1 or 0 for invalid?} &
\texttt{Consider returning 0 instead of -1 for invalid object IDs to maintain consistency with standard error handling conventions.} \\
\bottomrule

\end{tabularx}
\vspace{-1em}
\end{table*}

\begin{figure}
    \centering
    \includegraphics[width=1\linewidth]{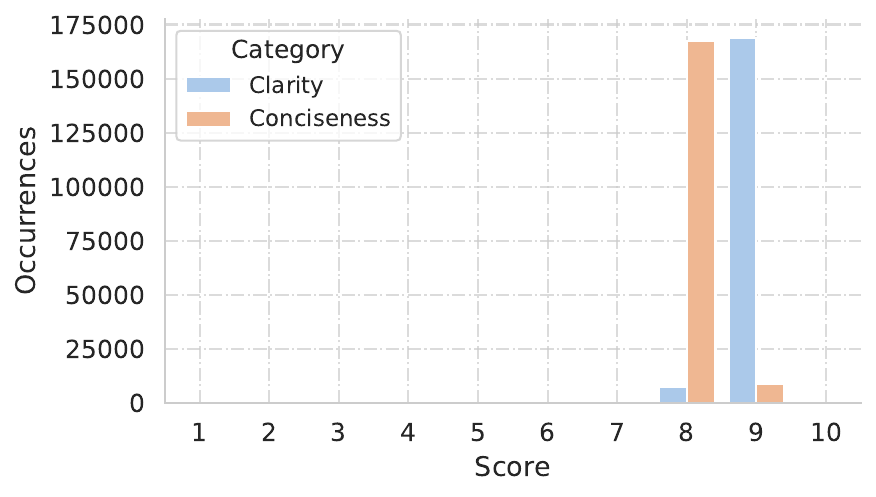}
    \caption{Distribution of the clarity and conciseness scoring criteria on the curated dataset.}
    \label{fig:curated_scoring_dist}
    \vspace{-1.5em}
\end{figure}

\begin{center}
\begin{tcolorbox}[colframe=purple!75!black, colback=purple!5!white, title=\textbf{Answer to RQ2}]
The results of the curation pipeline indicate that the curated dataset shifted towards clearer ($6.89 \rightarrow 8.96$), more actionable ($62.60\% \rightarrow 90.20\%$), more concise ($7.71 \rightarrow 8.05$), and more civil ($98.80\% \rightarrow 100\%$) review comments.
\end{tcolorbox}
\end{center}
\section{A Comparative Study of Initial and Curated Code Review Datasets}
\label{sec:analysis}

\subsection{Impact of Curated Reviews on Comment Generation}
\label{subsec:model_data}
In this section, we investigate the impact of curated reviews on automating the comment generation process. By comparing models trained on both original and curated comments, we aim to assess whether the reformulated reviews lead to more efficient automation of the comment generation task.

\paragraph{\textbf{Model and data selection}}
To ensure a fair comparison, we selected a subset of $20,000$ comments from both the original and curated datasets, such that each original review comment \( r_i \) from the original dataset is paired with its reformulated counterpart \( r'_i \) in the curated dataset. This selection strategy guarantees that any observed differences in model performance can be attributed to the quality of the data (\ie curation process) rather than differences in review content or model hyperparameters. We further split each subset into $75\%$ for training and $25\%$ for evaluation.

We selected \textit{DeepSeek-6.7B-Instruct} \cite{deepseek-coder}, an LLM tailored for code-related tasks. 
Given a code change, the model was tasked with generating either the original or the curated comment. To ensure consistency, we trained two separate models, one for each dataset version, using identical configurations.

\paragraph{\textbf{Experimental setup}}

This experiment aims to determine whether curated review comments improve the ability of LLMs to generate accurate review comments. For each dataset version (original and curated), we provided the model with code changes as input and tasked it with generating the corresponding review comment.

Each model was trained independently using the same configuration to ensure that observed performance differences could be attributed solely to the dataset quality, not to model or hyperparameter variations. The training was conducted on four \emph{NVIDIA RTX A5000 GPUs}, each with \emph{24GB} of memory. We used a batch size of $4$ and trained each model for $5$ epochs. To enable efficient, low-resource fine-tuning, we employed Low-Rank Adaptation (LoRA) \cite{hu2021lora}, a parameter-efficient fine-tuning technique, configured with settings of $r = 16$, $\alpha = 32$, and $dropout = 0.05$. LoRA operates by decomposing the weight updates of a neural network into low-rank matrices, significantly reducing the number of parameters that require updating during fine-tuning \cite{hu2021lora}, thus enhancing the overall efficiency of the training process.
LoRA has been widely used in prior work to fine-tune LLMs for software engineering tasks~\cite{lu2023llama, weyssow2023exploring, hou2023large, silva2023repairllama}

To evaluate the two produced models’ performance, we used the BLEU score \cite{papineni2002bleu}, a standard metric, widely used in the literature, that measures the precision of n-grams in the generated text relative to the ground truth. BLEU is well-suited for assessing the correctness of generated comments, with higher scores indicating greater accuracy with real output.

\paragraph{\textbf{Results}}
The results are presented in \Table{tab:com_results}. For the model trained on the original dataset, we obtained a BLEU score of $7.71$, whereas the model trained on the curated dataset achieved a BLEU score of $11.26$. This improvement suggests that the reformulated, curated comments are likely easier for the model to learn, potentially due to their enhanced clarity and structure.

These findings suggest that curated review comments provide clearer, more direct guidance, enabling the model to better capture the intended message and improving the quality of generated comments. The higher BLEU score with the curated dataset indicates that the curation process enhances the ability of models to generalize and learn producing more accurate review comments, thus facilitating a more efficient automation of the comment generation process.

% \begin{table}[!t]
% \centering
% \caption{Comparison of BLEU scores for DeepSeek-Coder-6.7B-Instruct trained on original and curated comments.}
% \label{tab:com_results}
% \begin{tabular}{@{}lcc@{}}
% \toprule
% \textbf{Dataset Version}       & \textbf{BLEU} \\ 
% \midrule
% Original Comments               & $7.71$ \\
% Curated Comments                & \textbf{$11.26$} \\ 
% \bottomrule
% \end{tabular}
% \vspace{-1.5em}
% \end{table}

\begin{table}[!t]
\centering
\caption{Comparison of BLEU scores for DeepSeek-Coder-6.7B-Instruct trained on original and curated comments.}
\label{tab:com_results}
\begin{tabular}{@{}lcc@{}}
\toprule
& \textbf{Original Comments} & \textbf{Curated Comments} \\ 
\midrule
\textbf{BLEU} & $7.71$ & $\textbf{11.26}$ \\ 
\bottomrule
\end{tabular}
\vspace{-1em}
\end{table}

The code change, shown in \Table{tab:comm_gen_examples}, is an example picked from our test set. The real comment in the original dataset lacks clarity, as it does not provide the reason behind the question. The model trained on the original dataset generates an incorrect review comment. In the curated dataset, the reformulated version of the real review comment is clearer, though less concise. The model trained on the curated dataset generates a comment that is accurate, closely matching the real and reformulated comments, and is even more concise than the reformulated version.

\begin{table*}[!htbp]
\centering
\caption{Example of real, reformulated, and generated review comments from the initial and curated datasets.}
\label{tab:comm_gen_examples}
\begin{tabularx}{1\linewidth}{X}
\toprule

\multicolumn{1}{c}{\textbf{Code change}} \\\midrule
\makebox[\linewidth]{\includegraphics[width=0.6\linewidth]{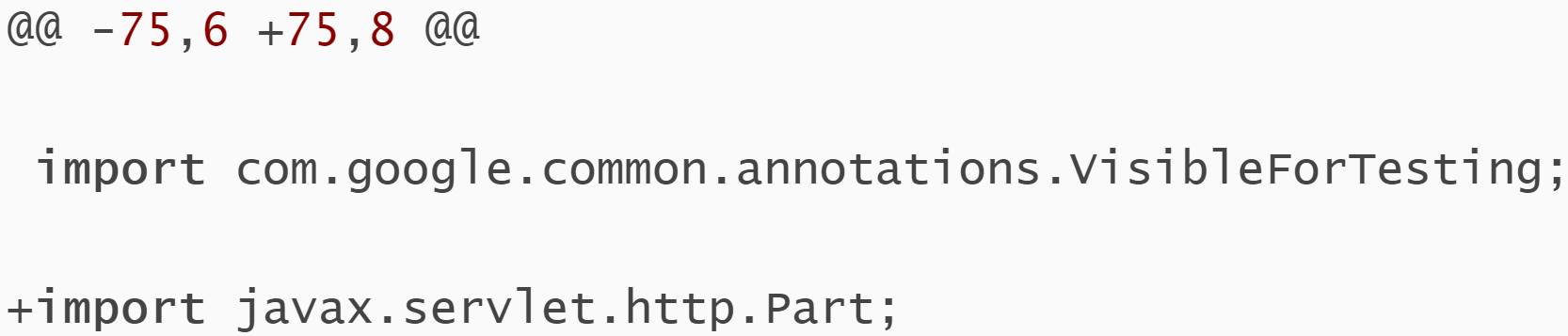}} 
\\\midrule
\multicolumn{1}{c}{\textbf{Original dataset}}\\\midrule

\underline{\textbf{Real review comment}} \texttt{why this import is required ?} \\

\underline{\textbf{Generated review comment}} \texttt{I think we should use `javax.servlet.http.Part`instead of \newline`org.apache.commons.fileupload.FileItem`.} \\ \midrule

\multicolumn{1}{c}{\textbf{Curated dataset}} \\\midrule

\underline{\textbf{Reformulated review comment}} \texttt{This import is not immediately clear in its necessity. Could you provide context\newline or explain why `javax.servlet.http.Part` is required for this functionality?} \\

\underline{\textbf{Generated review comment}} \texttt{Consider removing the import statement for `javax.servlet.http.Part` as it's not\newline used in the code.} \\ \bottomrule

\end{tabularx}
\vspace{-.5em}
\end{table*}

\begin{center}
\begin{tcolorbox}[colframe=orange!75!black, colback=orange!5!white, title=\textbf{Answer to RQ3}]
Curated review comments are more effective for training models on the comment generation task, as evidenced by higher BLEU ($7.71 \rightarrow 11.26$).
\end{tcolorbox}
\end{center}

\subsection{A Comparative Analysis on the Usefulness of Curated Comments for Code Refinement}

In this section, we aim to evaluate the usefulness of the curated comments compared to the original comments for code refinement.
We conduct a comparative study to assess which version of the comment—original or curated—guides the code refinement model to generate more accurate code changes.

\paragraph{\textbf{Model and data selection}}
% We select \textit{DeepSeek-Coder-6.7B-Instruct} \cite{deepseek-coder} to serve as our code refinement model. For evaluation, we randomly select a subset of our large datasets, choosing 20,000 samples from the full set of 176,613 for each version of the dataset (\ie original and curated datasets). To ensure a fair comparison, we maintain consistency in the selection by pairing each original review comment \( r_i \) from the original dataset with its corresponding reformulated comment \( r'_i \) in the cated dataset, where \( i \in [1, 20,000] \). Thus, each review pair \( (r_i, r'_i) \) represents the same code review context, allowing us to directly evaluate the impact of the reformulated comments compared to the original. For consistency, we use the same \textit{DeepSeek-Coder-6.7B-Instruct} model configuration on both datasets to ensure that any observed differences are attributable to the quality of the input data (\ie review comments) rather than model variations. 

We use \textit{DeepSeek-Coder-6.7B-Instruct} \cite{deepseek-coder} as a code refinement model, applying it to the same selected subset of $20,000$ samples from each dataset version (original and curated), as in the previous experiment on comment generation, as explained in \Sect{subsec:model_data}. To ensure a fair comparison, each original review comment \( r_i \) is paired with its reformulated counterpart \( r'_i \), preserving the same review context across both datasets. The model configurations remain identical for both datasets, to ensure that any observed differences are attributable to the quality of the input data (\ie review comments) rather than model variations.

\paragraph{\textbf{Experimental setup}}
For each dataset version, we provided the code refinement model with the original code diff, the old file, and the review comment (either original or curated) as context and prompted it to generate a code diff that accurately implements the specified changes.

We used the LLM directly for inference, as its extensive training on diverse code-related tasks equips it with the capabilities needed to effectively automate the code refinement task. The experiment was run twice, once with the original comments and once with the curated comments.

To evaluate the accuracy of the generated code diffs, we employed two evaluation metrics: 
\begin{itemize} 
    \item \textbf{CodeBLEU}: This metric measures the similarity of the generated code diff to the expected code diff, combining n-gram match, weighted n-gram match, AST match, and data-flow match scores \cite{ren2020codebleu}. 
    \item \textbf{Exact Match (EM)}: This metric calculates the number of generated code diffs that exactly match the expected code diff. 
\end{itemize}

Each experiment was conducted using identical model configurations for both dataset versions to ensure that any observed performance differences could be attributed solely to the quality of the review comments rather than model parameter variations.

\paragraph{\textbf{Results}}
The results, presented in \Table{tab:ref_results}, reveal substantial differences between the two dataset versions. Using the \textbf{original comments}, the model achieved a \emph{CodeBLEU} score of $0.36$ and an \emph{EM} of $408$. When utilizing the \textbf{curated comments}, the model performance improved significantly, reaching a \emph{CodeBLEU} score of $0.44$ and an \emph{EM} of $445$. These findings suggest that the curated comments offer more precise guidance, enabling the model to generate more accurate code changes that are closer to the ground truth.

This result indicates that the curated comments not only clarify the intended modifications but also reduce ambiguities in the model interpretation of the review instructions. The curated comments likely contain enhanced phrasing and structure that aid the model in better understanding and implementing the required code changes, thus improving the overall quality of the generated code diff.

\begin{table}[h!]
\centering
\caption{Comparison of DeepSeek-Coder-6.7B-Instruct's effectiveness for code refinement using the original and curated review comments.}
\label{tab:ref_results}
\begin{tabular}{@{}lcc@{}}
\toprule
\textbf{Dataset Version}       & \textbf{CodeBLEU} & \textbf{Exact Match} \\ \midrule
Original Comments & $0.36$ & $408$\\
Curated Comments  & $\textbf{0.44}$ & $\textbf{445}$\\ 
\bottomrule
\end{tabular}
\vspace{-.5em}
\end{table}

This experiment showed that curated review comments are more useful for the next task, as they provide better guidance for the code refinement task, leading to more accurate code changes. This is reflected by significant improvements in both \emph{CodeBLEU} and \emph{EM}.

\begin{center}
\begin{tcolorbox}[colframe=black!75!black, colback=black!5!white, title=\textbf{Answer to RQ4}]
    Curated comments demonstrate superior utility for guiding code refinement models, leading to notably higher CodeBLEU ($0.36 \rightarrow 0.44$) and Exact Match scores ($408 \rightarrow 445$).
\end{tcolorbox}
\end{center}

\section{Threats to Validity}
\label{sec:threats}

The evaluation results have shown that our proposed methodology is effective in curating review comments and improving the performance of downstream tasks (\ie comment generation and code refinement). However, certain threats may limit the validity of these evaluation results.

A primary threat pertains to the nature of the data, specifically the review comments, which may contain noise and potentially non-English or misspelled words. We have mitigated this by employing \emph{Llama-3.1-70B}, an LLM that utilizes Byte-Pair Encoding \cite{sennrich2015neural}, a subword-based tokenization algorithm. This algorithm breaks unseen words into several frequently seen sub-words that can be effectively processed by the model.

Another concern relates to the reliability of using an LLM as a judge, as LLM-generated judgments may not always match the accuracy of human evaluations. However, employing a bigger LLM, specifically \emph{Llama-3.1-70B}, helps mitigate this issue due to its advanced capabilities in producing accurate judgments. Prior research supports this hypothesis, showing that highly capable LLMs align closely with human assessments, often achieving agreement rates comparable to those between human evaluators \cite{zheng2024judging, li2023alpacaeval}. To further validate the reliability of LLM judgments in our study, we conducted a thorough sanity check on $100$ manually assessed samples. The obtained Cohen's kappa scores indicated strong agreement rates between human and LLM evaluations, reinforcing the reliability of \emph{Llama-3.1-70B} judgments.

\section{Related Work}
\label{sec:related}

To assist developers in code review, various techniques and tools have been proposed to automate review comment generation.
Early approaches relied on information retrieval methods. For instance, Hong \etal \cite{hong2022commentfinder} introduced CommentFinder, an approach that retrieves relevant past review comments for new code changes. Similarly, Gupta et al. \cite{gupta2018intelligent} developed DeepCodeReviewer (DCR), an LSTM-based model that predicts relevant reviews by ranking them based on code similarity. Although these methods effectively recommend existing comments, their limitation lies in the inability to generate new comments for unseen code.

More advanced solutions leverage language models. Tufano \etal \cite{tufan2021towards, tufano2022using} used the T5 transformer, pre-trained with a masked language modeling task, and fine-tuned to generate review comments for Java code. Building on this, Li \etal \cite{li2022automating} introduced a CodeT5 model pre-trained on tasks specifically designed for code review, such as quality estimation and comment generation. This demonstrated significant advancements in handling multilingual datasets and downstream tasks, including comment generation and code refinement.

Recent work aimed to enhance the performance of the finetuning-based approaches. Sghaier \etal \cite{sghaier2024improving} proposed DISCOREV, an approach that incorporates cross-task knowledge distillation, connecting quality estimation, comment generation, and code refinement. This technique uses a cascade of models where each task informs the fine-tuning of the next, showing improvements over previous models.

Although these approaches have shown promising results, none of the existing works in the literature examined the quality of code review datasets or implemented preprocessing techniques to curate the data. Instead, most efforts focused on the fine-tuning phase of pre-trained language models. This leaves a significant gap in addressing the limitations of existing raw datasets. Our work proposes curating a code review dataset to improve the automation of code review tasks, overcoming the challenges posed by noisy and unrefined data.

\section{Conclusion}
\label{sec:conclusion}

In this work, we present a methodology for enhancing the quality of code review datasets. We first propose an evaluation schema for review comments and conduct an empirical study to assess the dataset quality, identifying key issues and areas for improvement. Building on these findings, we introduce a curation process aimed at reformulating review comments to improve their quality, as measured by several criteria. We then perform a comparative analysis between the original and curated datasets. We evaluate the impact of these datasets on automating downstream code review tasks, specifically comment generation and code refinement.

Our findings open up numerous research opportunities. The insights from our empirical study could guide work on predicting the type of issue in a code change, balancing training datasets across different categories, and filtering datasets based on scoring criteria to produce more robust and unbiased models. Additionally, our curated dataset offers a valuable baseline for future research in code review, particularly in comment generation, where its improved quality will enable the development of models that generate relevant, clear, concise, and civil feedback.

In future work, we plan to incorporate multiple LLMs as juries to enhance dataset curation, fostering a diverse set of review comments. We also aim to explore prioritization strategies for review comments, helping reviewers efficiently prioritize pull requests based on urgency and relevance.

\section*{Data Availability}
We publicly release all the code \cite{github_replication}, models, data, and results \cite{zenodo_data} of our experiments.

\bibliographystyle{IEEEtran}
\bibliography{references}

\end{document}